\documentclass[10pt]{article}
\usepackage{sao2}
\usepackage{psfig,epsf}

\issue{2011, 55, 302--309}

\begin{document}
\markboth{M.L.~Khabibullina and O.V.~Verkhodanov }
{THE ESTIMATION OF BLACK-HOLE MASSES IN DISTANT RADIO GALAXIES}
\title{The Estimation of Black-Hole Masses in Distant Radio Galaxies}
\author{M.L.~Khabibullina\inst{a},
O.V.~Verkhodanov\inst{a}
}
\institute{
$^a$\saoname}

\date{2010}{October 22, 2010}
\maketitle

\begin{abstract}
We have estimated the masses of the central supermassive black holes of 2442
radio galaxies from a catalog compiled using data from the NED, SDSS, and
CATS databases. Mass estimates based on optical photometry and radio data are
compared. Relationships between the mass of the central black hole
$M_p^{bh}$
and
the redshift $z_p$ are constructed for both wavelength ranges. The distribution
of the galaxies in these diagrams and systematic e ects influencing
estimation of the black-hole parameters are discussed. Upper-envelope cubic
regression fits are obtained using the maximum estimates of the black-hole
masses. The
optical and radio upper envelopes show similar behavior, and have very
similar peaks in position, $z_p \simeq 1.9$ and amplitude, $\log M_p^{bh}$=9.4.
This is consistent with a model in which the growth of the supermassive
black holes is self-regulating, with this redshift corresponding to the
epoch when the accretion-flow phase begins to end and the nuclear activity
falls off.

\end{abstract}

\maketitle

\section{INTRODUCTION}
One of the central questions in studies of distant radio galaxies (redshifts
$z > 0.3$) is the origin and growth rate of supermassive black holes (SMBHs) in
 their nuclei. The discovery of radio galaxies and quasars at high redshifts
has raised the need to explain the rapid formation of SMBHs in early epochs
of the evolution of the Universe. For example, radio galaxies have been
detected at z = 5.19 (TN J0924-2201) \cite{breugel}, when the age of the Universe
was only $t\sim1.1\times10^9$\ yrs (in a cosmology with
$\Omega_m=0.27$, $\Omega_\Lambda=0.73$, H\_0 = 70 km s-1Mpc-1), and z = 4.515
\cite{kopylov}, when $t\sim1.3\times10^9$\
yrs. Quasars have been found in the Sloan Digital Sky Survey (SDSS) at
 redshifts of z = 6.42 (J114816.64+525150.3), z = 6.23 (J104845.05+463718.3),
 and z = 6.05 (J163033.90+401209.6) \cite{fan}, when the age of the Universe was
$t\sim$850-900 million years. Willot et al. \cite{willot} used the width of the emission
lines of the quasar J114816.64+525150.3 at z = 6.42 to estimate the
mass of its SMBH to be $M_{bh} = 3\times10^9$ $M_\odot$. The rapid (over 800 million years)
formation of SMBHs
in a $\Lambda$CDM cosmology requires special models for their growth rates.
As was noted by Loeb \cite{loeb}, if the formation of massive black holes is
described by a heirarchical scheme similar to that for the formation of their
parent galaxies, this process must be rapid and e cient. Models with
primordial massive black holes that formed after the Big Bang but before the
formation of galaxies are often invoked to explain the early existence of
SMBHs.

It is believed that SMBHs play a central role in the formation and evolution
 of massive galaxies, and are also a key component in the development of
nuclear activity. However, it remains unclear how galaxies and their central
 SMBHs are related to the formation of observed structures \cite{silk}-\cite{adams}. An
important result obtained from measurements of black-hole masses in nearby
galaxies is the existence of an appreciable correlation between the mass of
 the central black hole and the mass of the bulge of the host galaxy, which
 gives rise to correlations between the black-hole mass and the luminosity
of the bulge \cite{magorian},\cite{kormendy1} and between the black-hole mass and the stellar velocity
 dispersion $\sigma$ (e.g. \cite{ferrarese}-\cite{kormendy2}. This result is based on a small number of nearby
 galaxies for which direct black-hole mass measurements are available ($\sim$30),
and there is appreciable scatter in the relation ($\sim$0.4 in the logarithm of
he black-hole mass). Nevertheless, these empirical relations provide a new
 tool for estimating black-hole masses in
various types of active galactic nuclei (AGNs), when the bulge luminosity
and/or stellar-velocity distribution is known \cite{mcLure1}-\cite{betton}. The existence of
these correlations can provide constraints on the evolution of AGNs, and
 can be used to test whether AGNs precisely follow these relations or not.

Franceschini et al. \cite{franceschini} established the existence of a correlation
between the radio power of AGNs and the masses of their black holes. A number
 of studies have also demonstrated relationships between black-hole masses
and radio powers for samples of
nearby galaxies, but these results differ from one another \cite{yi}-\cite{matteo}.
Discussions of this relation were also extended to more active galaxies,
 such as Seyfert galaxies and quasars (see, for example, \cite{laor},\cite{nelson}-\cite{gu}).

A number of studies have used data from the SDSS \cite{schneider} for quasars to estimate
 the growth rates of their SMBHs. Kelly et al. \cite{kelly} constructed the black-hole
 mass function for a sample of 9886 broadline quasars with redshifts
1 < z < 4.5. Their results support the "downsizing" effect for black holes in
 broad-line quasars, whose peak number density shifts
toward high z as the black-hole mass is increased\footnote{\tt The ``Downsizing'' scenario
scenario, which can explain a mass dependence of the evolutionary history in
 which less massive elliptical galaxies have longer star-formation histories
 than more massive elliptical galaxies.}.
This peak occurs at
z$\sim$2. Moreover, Kelly et al. \cite{kelly} estimated the completeness of the SDSS
 sample as a function of black-hole mass and the ratio of the luminosity
of a quasar to the Eddington luminosity, $L/L_{Edd}$, and found that the sample
of black-hole masses
$\le 10^9$ í$_{\odot}$ and $L/L_{Edd}\le0.5$ is very incomplete at redshifts
z > 1. According to their model,
they estimate the lifetime of the broad-line quasar stage to be approximately
 150 million years at z = 1 for a black-hole mass of $\sim 10^9$ $M_\odot$. Shen et al.
\cite{shen} constructed a sample of 105783 quasars based on the SDSS, and used
various properties (emission in H$\alpha$, H$\beta$, MgII, CIV; radio data; broad
absorption lines) to estimate their black-hole masses. They also obtained a
logarithmic dependence between the black-hole masses and the luminosities and
 widths of the lines used.

Here, we verify two dependences of the black-hole masses on the radio and
 optical luminosities that are used in the literature in studies of radio
 galaxies. We used data from the SDSS, NED, and CATS databases to construct
 a sample of distant (z > 0.3) radio galaxies, for which we derived and
compared these dependences for the black-hole mass. We calculated the
luminosities using optical data in the R filter and 5-GHz radio flux
densities. We assumed a $\Lambda$CDM cosmology with H0 = 71 km s-1Mpc-1,
${\Omega_{M}}$= 0.27,
 and ${\Omega_{\Lambda}}$ = 0.73. We used a catalog of distant radio galaxies with
spectroscopic redshifts z > 0.3, which were selected based on data from the
 largest available databases \cite{kha1}-\cite{ver1}.

\section{STUDIED SAMPLE AND MASS ESTIMATES}
\subsection{Input Data}

A sample of radio galaxies with z > 0.3 \cite{kha1}-\cite{ver1} was constructed using the
 NED ({\tt http://nedwww.ipac.caltech.edu}),  CATS  ({\tt http://cats.sao.ru}) \cite{ver2},\cite{ver3},
and SDSS ({\tt http://www.sdss.org}) \cite{gu} databases. This sample can be used to
carry out various statistical and cosmological tests requiring a
comparatively large number of objects of a single type \cite{ver3}-\cite{ver5}. The NED
 database was used to construct the initial list of sources, selecting
objects with specified parameters, most importantly with redshifts z > 0.3
 and morphologies corresponding to radio galaxies. The initial catalog
contained 3364 objects. This galaxy sample was contaminated by objects with
 incomplete information and objects with other properties. Therefore, we paid
 special attention to eliminating unwanted sources from the initial sample,
 including (1) galaxies with redshifts determined using photometric methods
 and (2) galaxies with quasar-like properties according to data from the
literature. The final catalog contains 2442 sources with spectroscopic
redshifts, photometric data and radio flux densities, radio sizes, and radio
 spectral indices, calculated via cross identification with other radio
catalogs in the CATS database at frequencies from 30 GHz to 325 MHz.

\subsection{Estimating the SMBH Masses. Optical Data}

To estimate the SMBH masss, we used the relationship between the black-hole
 masses and R absolute magnitudes ($M_{bh} - M_{\rm R}$) from \cite{kormendy2}, which was obtained
 via an analysis of the magnitudes and stellar velocities in AGNs:
$$
log(\frac{M_{bh}}{M_{\odot}}) = -0.50(\pm 0.02){M_{\rm R}} - 2.27(\pm 0.48)\,,
$$
where ${M_{\rm R}}$ is the absolute R magnitude of the
 bulge, assuming that the bulge of an elliptical galaxy essentially represents
the entire galaxy.

Since R observations are not available for all the catalog radio galaxies,
 we estimated the R luminosities of a number of objects based on data in
other filters using the database of galactic spectral-energy distributions
(SEDs) \cite{ver6}. Photometric data were used to estimate the ages of the radio
galaxies, assuming that their SEDs corresponded to the SEDs of elliptical
 galaxies. We determined the R magnitude using the redshift and the age of
 the corresponding model SED track. We used the GISSEL model data \cite{bruzual},
which contains synthetic spectra of elliptical galaxies for ages from 200
 million years to 14 billion years, as the basic library of tracks. This
library can be accessed both in the HyperZ package \cite{bolzonella} and at the site
{\tt http://sed.sao.ru}
\cite{ver6}. The distribution of estimated and measured R
magnitudes for the radio-galaxy sample is shown in Fig. 1.

\begin{figure*}
\centerline{
\hbox{
\psfig{figure=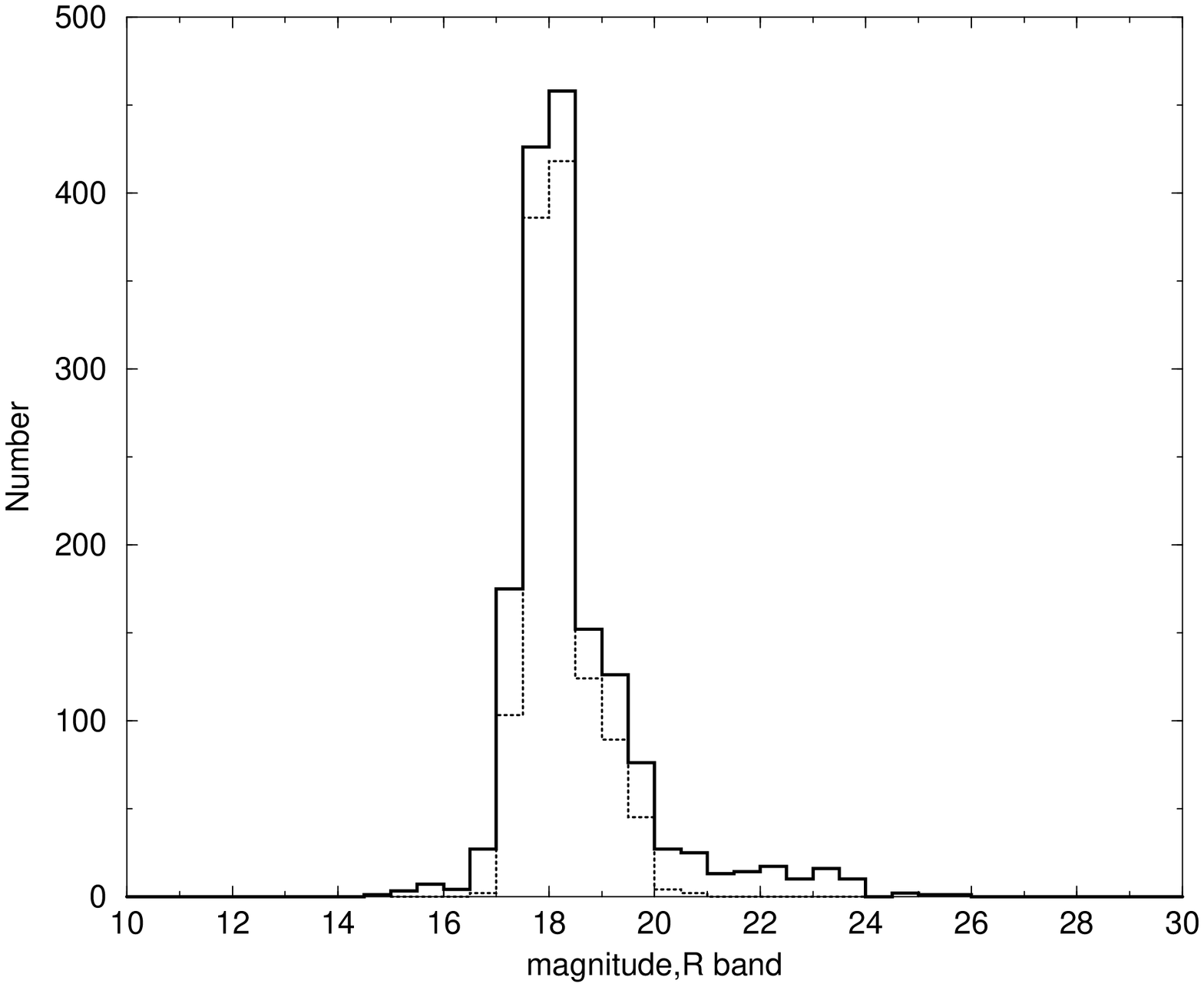,width=7cm,angle=-90}
\psfig{figure=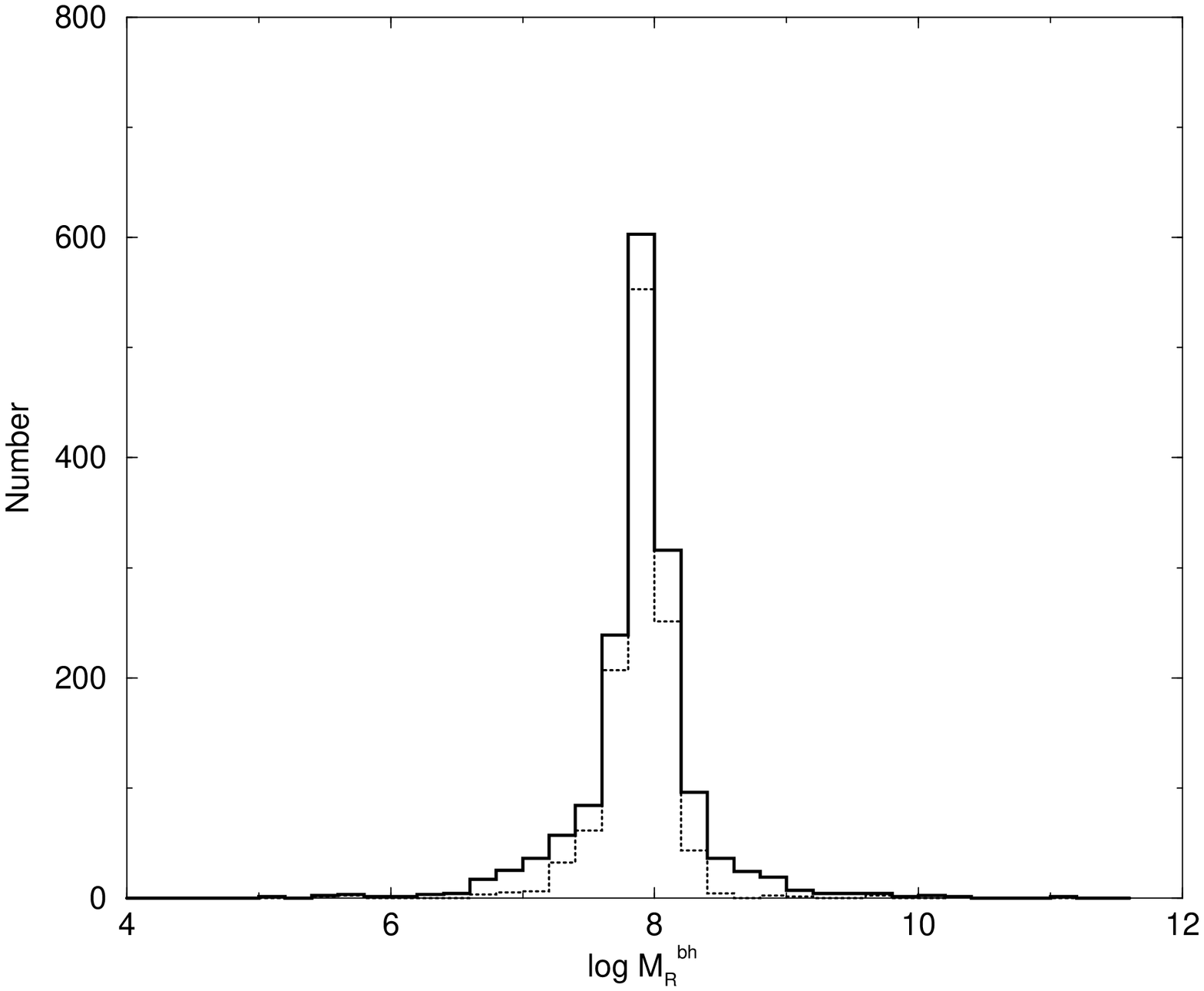,width=7cm,angle=-90}
}}
\caption{Left: distribution of R magnitudes for the sample of radio galaxies.
 Right: distribution of black-hole masses obtained from R data for the sample
 of radio galaxies. The SDSS subsample of radio galaxies is shown by the
dotted line.}
\label{f1}
\end{figure*}

The luminosity was calculated using the formula \cite{lang} (see also \cite{sahni})
$$
L =   4 \pi F {d_{L}^{2}}(z),
$$
where F is the measured flux,
$
{d_{L}}(z) = (1 + z) \int\limits_{0}^z \frac{dz^{'}}{H({z^{'}})}\,,
$
is the  luminosity  distance, and
$
H(z) = {H_0}{[{\Omega_R}{(1+z)^{4}}+{\Omega_m}{(1+z)^{3}}-({\Omega_0}-
1){(1+z)^{2}}+{\Omega_{\Lambda}}]^{1/2}} \,.
$
is the Hubble parameter. The resulting luminosity (more precisely, absolute
 magnitude) was used to estimate the masses of the central black holes of
the galaxies. The distribution of estimated black-hole masses is shown in
Fig. 1.

\subsection{Estimating the SMBH Masses. Radio Data}

To calculate the estimated black-hole masses based on radio data, we used
 the relationship between the radio luminosity and black-hole mass
($P_{5\,GHz} - M_{bh}$) from \cite{franceschini}:
$$
P_{5 GHz} \propto  {\Big(\frac{M_{bh}}{M_{\odot}}\Big)^{2.2\div3.0}}\,,
$$
where  ${P_{5\,GHz}}$ is the total power at 5\,GHz.
Our calculations used the formula
$$
P_{5 GHz} \propto  {\Big(\frac{M_{bh}}{M_{\odot}}\Big)^{3.0}}\,.
$$
We calculated the radio power using the standard formula \cite{lang}
$$
{P_{\nu}} = 4 \pi {d_L^{2}} S\Big(\frac{\nu}{(1+z)}\Big) {(1+z)^{-1}}\,,
$$
where $S\Big(\frac{\nu}{(1+z)}\Big)$ is the source flux density at frequency
$\nu$ taking into account the redshift and $d_L$ is
the photometric distance. The value $S\Big(\frac{\nu}{(1+z)}\Big)$ was
determined from the continuum radio spectrum of
each source, constructed via fitting with a standard set of functions. We
used the spg program \cite{ver7} of the RATAN-600 continuum data-reduction system
for this purpose. For radio galaxies with measurements at only one frequency,
we took the spectral index to be equal to the median value for our sample
\cite{kha1} at
1.4 GHz ($\alpha_{med}= -0.63$). The distribution of the obtained radio
power at 5 GHz is shown in Fig. 2.

\begin{figure*}
\centerline{
\hbox{
\psfig{figure=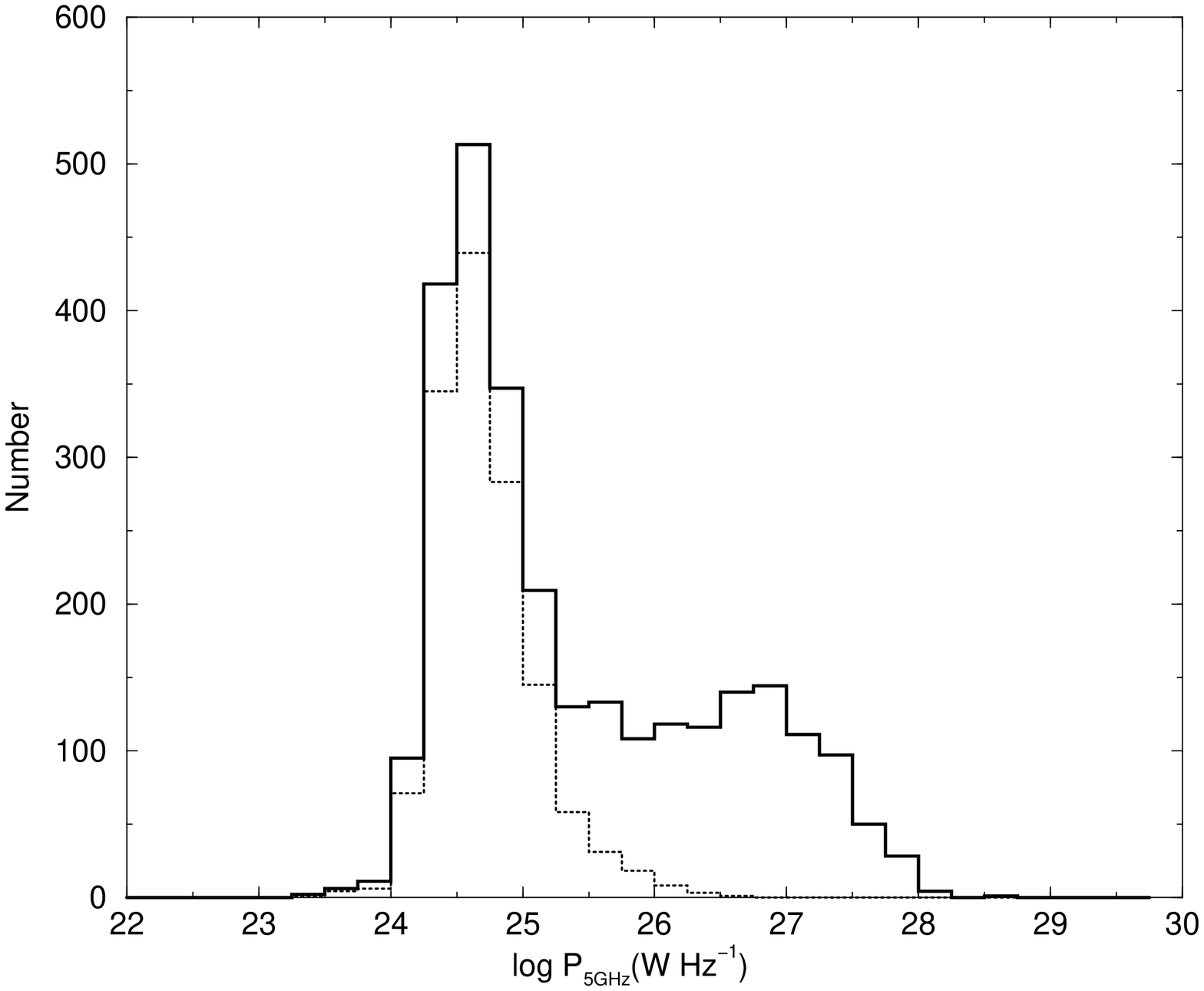,width=7cm,angle=-90}
\psfig{figure=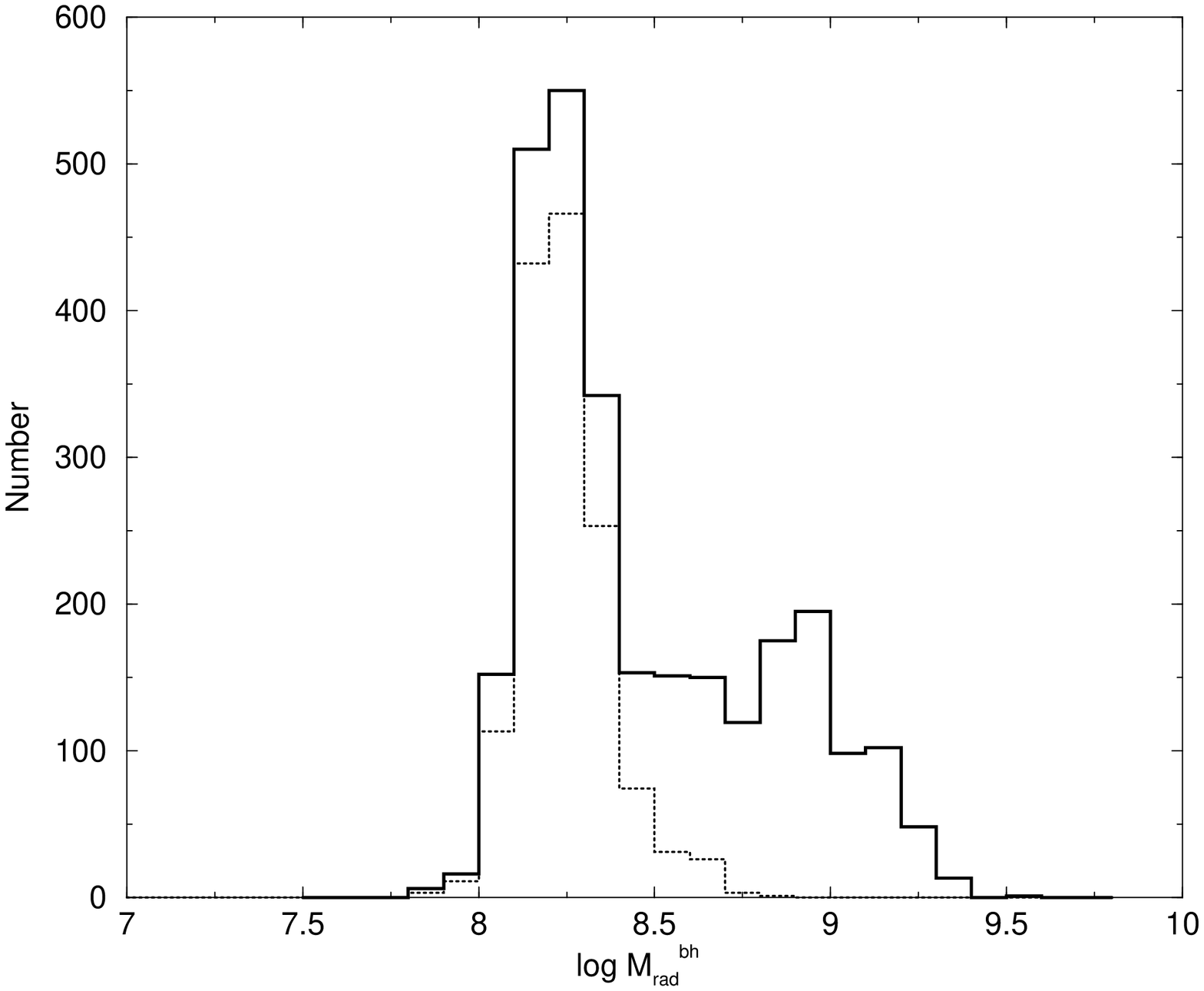,width=7cm,angle=-90}
}}
\caption{Left: distribution of the 5-GHz radio power for the radio-galaxy
sample. Right: distribution of the black-hole masses obtained from the total
 radio power (luminosity) at 5 GHz. The SDSS subsample of radio galaxies is
shown by the dotted line.}
\label{f3}
\end{figure*}

\section{DISCUSSION}

A comparison of the masses estimated using radio and optical data shows a
clear difference in the shapes of the mass distributions (Figs. 1 and 2).
Moreover, in contrast to the distribution of optical mass
estimates, the distribution of radio mass estimates is two-peaked. This could
 result from evolutionary manifestations in the radio flux at di erent
redshifts at which the radio sources are ignited and then quiet down. Then,
we may obtain di erent estimates for the mass of the "central engine" based
on the radio flux, corresponding to di erent stages of activity. The higher
peak in the distribution in Fig. 2 corresponds to closer objects with lower
masses, observed in the SDSS. Note that, in spite of the two-peaked form of
the mass distribution, the radio data provide a smaller scatter in the
redshift distribution (Fig. 3).
We should note that nonlinearity of ``radio luminosity --- optical
luminosity'' relation having the underlying relation of radio luminosity
and black hole mass was demonstrated in 1960-s by Pskovskii \cite{pskovskii}
and corrected by Iskudarian and Pariiskii \cite{iskudarian}.

\begin{figure*}[!th]
\centerline{
\vbox{
\hbox{
\psfig{figure=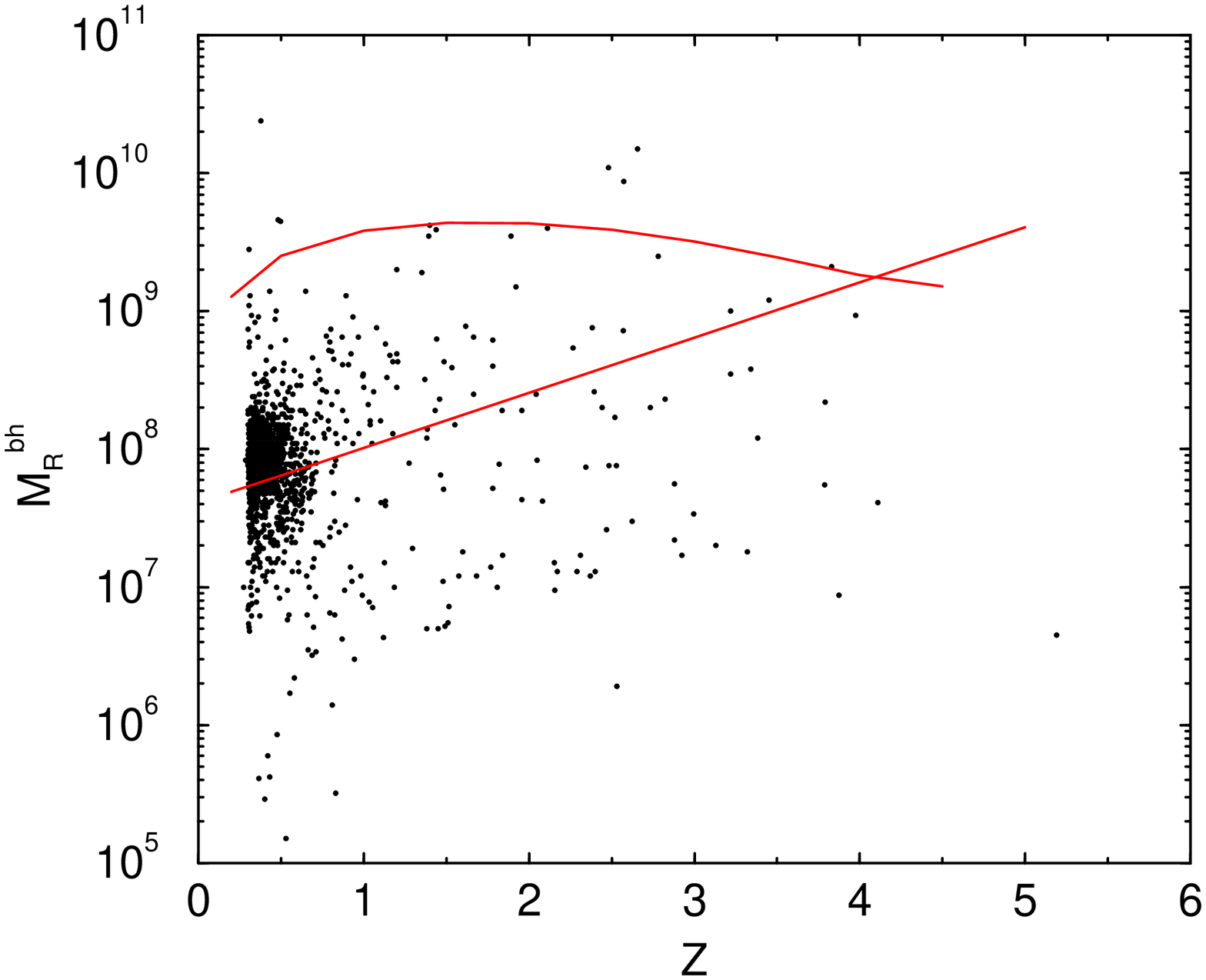,width=9cm,angle=-90}
\psfig{figure=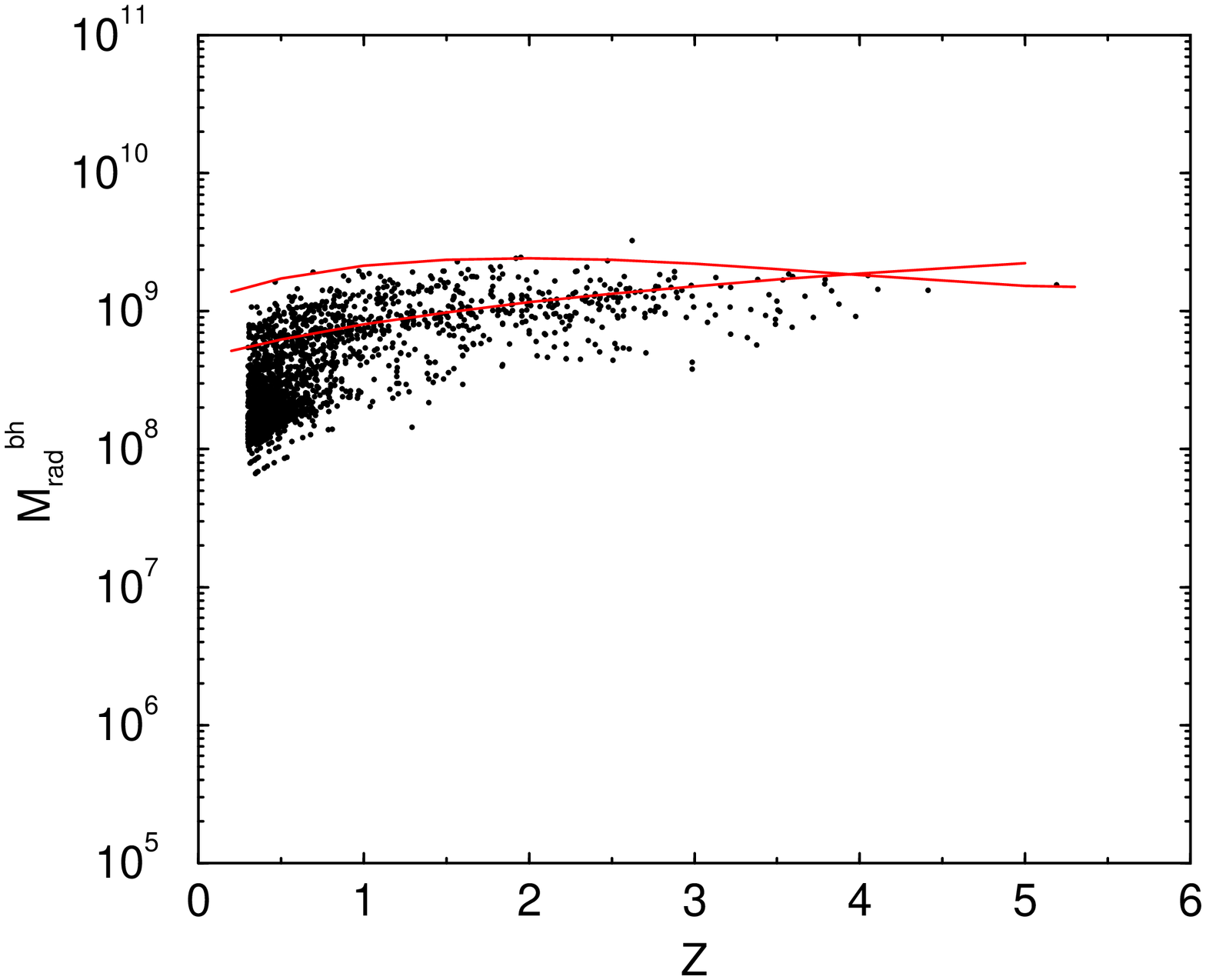,width=9cm,angle=-90}
}
}}
\caption{Plots of black-hole mass versus redshift for the radio galaxies,
based on R optical data (left) and 5-GHz radio data (right). Regression fits
were constructed for the mean and maximum black-hole masses in intervals
$\Delta z = 0.5$.}
\label{f6}
\end{figure*}

While the optical black-hole mass estimates can tentatively be divided into
two subgroups--strongly clustered at z < 0.7 and weakly clustered at
z > 0.7--this is not the case for the radio mass estimates. The range of
black-hole masses derived using the radio data is two orders of magnitude
 smaller than that for the optically derived masses.
The mean radio masses in individual redshift intervals ($\Delta z=0.5$)
are higher than the corresponding optical masses. Although the radio
estimates also display clustering at z < 0.7 for the first subsample, the
objects of the second subsample are grouped along the line for the median
regression fit.

Figure 3 shows the least-squares regression fits and upper envelopes for
both types of data. For the optical data (Fig. 3, left), the fit for the
median values on the log MRbh-z diagram\footnote{\tt Here and below, the
black-hole masses are expressed in solar masses, M$\odot$} in bins $\Delta z = 0.5$
 yields a growing
function satisfying the linear relation
$\log {M_{R}^{bh}} = 0.387(\pm0.134) \cdot z + 7.665(\pm0.309)$,
where $M_{R}^{bh}$ is the black-hole mass estimated using the R optical data. The
upper envelope is described by a cubic function:
$\log {M_{R}^{bh}} = 9.058+0.783\cdot z-0.303\cdot{z^{2}}+0.030\cdot{z^{3}}$.
Similarly, the linear fit for the radio data (Fig. 3, right) has the form
$\log {M_{rad}^{bh}} =0.126(\pm0.027)\cdot z + 8.793(\pm0.073)$,
the upper envelope the form
$\log {M_{rad}^{bh}} = 9.084+0.347\cdot z
		-0.120\cdot{z^{2}}+0.011\cdot{z^{3}}$.
In spite of the difference
in the dispersion of the mass estimates, the posi-tions and amplitudes of the
 maxima of both upper envelopes are similar: the peak is at $z_p = 1.78$ and
$\log M_p^{bh}$ = 9.67 for the optical data, and at $z_p = 1.92$
and $\log M_p^{bh}$=9.38 for the radio data.

The location of the maxima in the redshift range 1.5 < z < 2 (with
similar values for both the optical and radio estimates) could be associated
 with some sort of systematic e ect reflecting real physical pro-cesses. This
 range of z corresponds to an epoch of massive "ignition" of radio sources as
 a result of mergers between galaxies in galaxy clusters. In this case, due
to selection effects, we will most likely detect objects with the maximum
luminosity, which
means those with the maximum estimated black-hole masses, in this range.

Note that this result obtained for our list of radio galaxies is consistent
with the results of \cite{kelly}, where a sample of broad-line quasars from the SDSS
 was considered. Kelly et al. \cite{kelly} found that the peak num-ber density of
SMBHs in broad-line quasars occurs at z ~ 2. The location of the
mass-distribution peak at $z\sim2$ is consistent with a self-regulating growth
 model for the SMBHs, with this being the epoch when the end of the
accretion-flow phase begins and the nuclear activity starts to fall off. Note
 that our estimates and the estimate of \cite{kelly} are also consistent with
self-regulation of the rate of formation of galaxies in halos of dark matter
via the interaction of the jets from the central object with the
intergalactic gas \cite{rawlings}.

A growth in the black-hole mass with increasing redshift is visible in both
diagrams, demonstrated by the linear fits. The uncertainties in the fit
parameters characterize the statistical scatter in the masses (and
luminosities) derived using the optical and radio data. The large scatter for
 the optical estimates, in particular the many deviations from the maximum
values, indicates small ratios of the optical to the Eddington luminosities,
 and may suggest that some Seyfert 2 galaxies have been included in our
sample. The linear fit based on the median masses may reflect selection
effects in the detection of distant objects, and the region of intersection
with the upper envelope
($z_c=4.14,  \log(M_c)=9.27$ for the optical and
$z_c=3.57, \log(M_c)=9.25$ for the radio) shows an upper limit, beyond
which this effect will dominate. Note, however, that only a few radio galaxies
 have been detected in this region \cite{kopylov}.

The $M^{bh}_{opt} - M^{bh}_{rad}$ relation indicates the region in which the
two methods used to estimate the masses of the SMBHs are statistically
applicable, and shows the amount of scatter obtained using these methods
(Fig. 4). Three regions of clustering can be distin-guished on this plot, for
 two of which we obtained
fits in the form
$\log {M_{R}^{bh}} = a\cdot \log {M_{rad}^{bh}}+b$
The parameters for the first group of clustered points are
$a=-0.036\pm 0.026$, $b=8.122\pm 0.219$, and those for the second group of
points $a=1.211\pm 0.029$, $b=-2.329\pm 0.261$. The data for radio galaxies
from different redshift ranges are shown by different symbols: $0.3\le z<0.7$
(points),
$0.7\le z<1.5$,(crosses), and
$1.5\le z$ (triangles). The high concentration of points in the left part of
 the diagram is due to the large number of comparatively nearby radio
galaxies in the SDSS, for which relatively similar black-hole masses were
obtained.

\begin{figure}
\centerline{
\hbox{
\psfig{figure=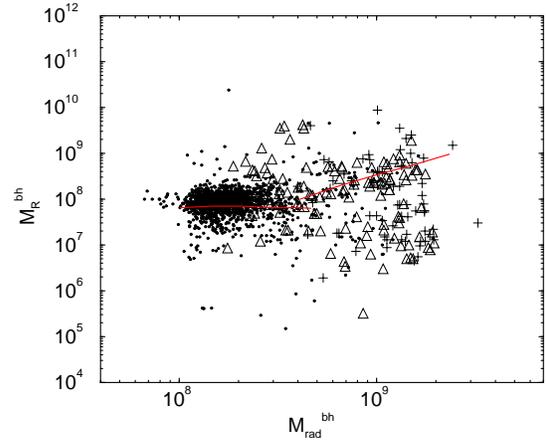,width=7cm,angle=-90}
}}
\caption{ Plot of $M^{bh}_{opt}$ versus $M^{bh}_{rad}$ for R-band and 5-GHz data. Regression
fits were obtained for two regions where the points are concentrated. Data
for radio galaxies from di erent redshift ranges are denoted with different
symbols: $0.3\le z<0.7$(points),
$0.7\le z<1.5$ (crosses), and
$1.5\le z$(triangles).}
\label{f5}
\end{figure}

Figure 4, and also Figure 1, show that the mean estimated masses for the
"central engines" for the SDSS subsample of galaxies are $\sim10^8$ $M_\odot$ based on
the optical data and from 10$^8$ to $10^9$ $M_\odot$ based on the radio data. In
general, nearly all the radio galaxies in the sample have
$M^{bh}_{opt} \le M^{bh}_{rad}$. For a second group, many of which are distant radio galaxies with
$z\ge1.5$ we observe a correlation between the two mass estimates for the
central objects. The objects in the lower right part of Fig. 4 have low
optical and high radio mass estimates. These low estimates from
the region of the lower envelope on the left Mbh(z) diagram in Fig. 3 and the
 corresponding estimates in the right diagram may be associated with
non-elliptical subclasses of radio galaxies.

We can distinguish several selection e ects that are inevitable when
estimating the central black-hole masses, even when care is taken in
selecting galaxies according to their optical and radio properties. Their
manifestation can explain differences between results obtained using different
 methods to determine the central black-hole masses. These selection effects
 include the following:

1)more distant objects tend to be more powerful; i.e., a sample of distant
objects contains, on average, galaxies with higher black-hole masses;

2)although the sample has been cleaned of objects with the properties of
 quasars (such as broad H$\alpha$ emission), some such objects with relatively weak
lines may remain in the subsample of comparatively nearby (z < 0.7) sources;

3)the SDSS objects in our catalog should not include Seyfert 2 galaxies with
 narrow lines; in spite of their radio emission, these galaxies have low
ratios of their bolometric to Eddington luminosities (see, e.g., \cite{lu}), and
their luminosities should thus place them in the lower part of the
diagram (Fig. 2);

4)the presence of evolutionary effects that are strongly manifest in the
radio, such as the fact that some radio galaxies may be observed in early
 phases of their activity, and that galaxies with central black holes of the
same mass may have radio fluxes differing by an order of magnitude;

5)some comparatively nearby objects whose peak activity in the radio has
passed have black-hole mass estimates derived from velocity dispersions
calibrated
based on the radio fluxes; in this case,the use of dependences for more
distant galaxies can lead to overestimation of the black-hole masses.

Nevertheless, the similarity of the behaviors of the mass upper envelopes
(Fig. 3) suggests that these estimates are trustworthy for objects with the
highest luminosities (close to the Eddington values). The comparatively small
 scatter of the radio masses for various z makes it possible to use this
diagram to estimate masses with a mean accuracy of less than an order of
magnitude, or even half an order of magnitude at z > 1.5. This is confirmed
by the plot in Fig. 4,
where a correlation between the radio and optical masses is observed for most radio galaxies with z >
$z>1.5$.

We used the other parameters in our catalog to construct plots of the
black-hole mass versus radio spectral index (Fig. 5) and versus the apparent
 size of the radio source (Fig. 6) for a single type of galaxies with active
nuclei (i.e., radio galaxies). The former plot shows a correlation between
the radio mass estimates and the radio spectral index
$$
\alpha = 2.73(\pm1.56) - 0.18(\pm0.08) \cdot \log (M_{bh})\,,
$$
There is a tendency for more powerful radio galaxies, which presumably have
 higher nuclear activiity, and possibly higher masses for their central
engines, to have steeper spectra (i.e., smaller spectral indices). Of course,
 it cannot be ruled out that some selection e ect is hidden via the redshift
 dependence of the spectral index $\alpha(z)$ found in \cite{ver1}. Distant radio galaxies
are often selected in surveys based on their spectral indices, and distant
sources tend to be more powerful; therefore, this e ect should make some
contribution to the detected dependence. However, the physical properties of
 the objects themselves at various z probably dominate in the $\alpha(z)$ relation.

\begin{figure}[!th]
\centerline{
\psfig{figure=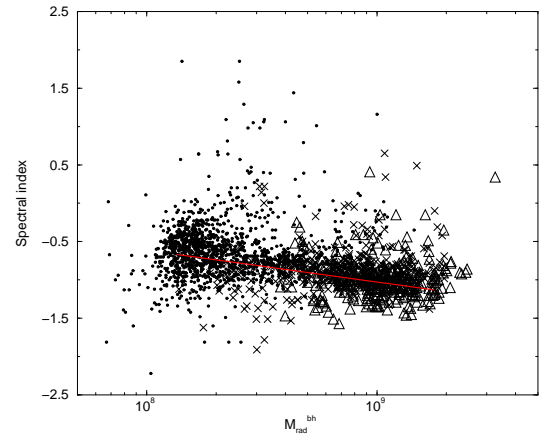,width=7cm,angle=-90}
}
\caption{Plot of black-hole mass versus radio spectral index. The masses
were estimated using the radio data. Notation is the same as in Fig. 4.}
\label{f5}
\end{figure}

\begin{figure}[!th]
\centerline{
\psfig{figure=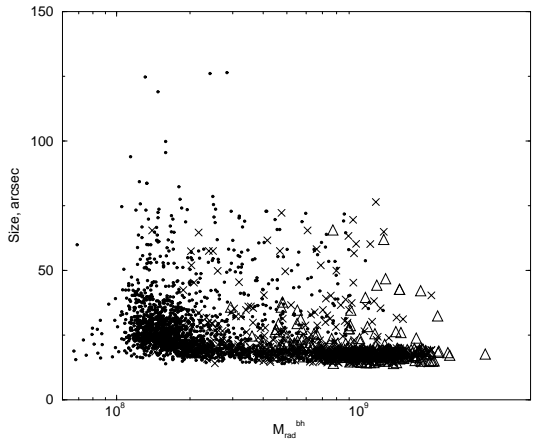,width=7cm,angle=-90}
}
\caption{Masses estimated using the radio data. The apparent sizes of the
galaxies were taken from the NVSS catalog \cite{rawlings}. Notation is the same as in
 Fig. 4.}
\label{f6}
\end{figure}

Figure 6 shows the distribution of the angular sizes of radio galaxies
according to the NVSS data \cite{condon} as a function of their radio black-hole mass
 estimates. Most of the objects distributed along the horizontal axis have
small angular sizes (<25$\arcsec$), determined by the resolution of the NVSS
($\sim$45$\arcsec$), and most of these are SDSS objects with redshifts z < 0.7. Note that
 the central objects of the most extended objects (> 1$\arcmin$) cover the entire
range of masses, and are encountered in all three redshift intervals we have
considered.

\section{CONCLUSION}

We have carried out a comparative analysis of estimates of the
central black-hole masses of 2442 radio galaxies with z > 0.3 \cite{kha1}-\cite{kha3},
derived from relations between the black-hole mass and the R
luminosity \cite{mcLure1} and between the black-hole mass and
the radio power \cite{franceschini}. Appreciable differences between
these  two  estimates are observed for many of the
radio galaxies. We have discussed various systematic
effects associated with observational selection effects
and the evolutionary properties of the radio galaxies
that could lead to these differences.  However, a diagram
of $M_{R}^{bh}$  versus $M_{rad}^{bh}$  (Fig. 4) reveals a region
where these two mass estimates are correlated. This
zone is formed primarily by the distant radio galaxies
in our sample. Moreover, the upper envelopes constructed
using the maxima of the two mass estimates
(Fig. 3) show similar behavior and have very similar
positions ($z_p \simeq 1.9$) and amplitudes ($\log M_p^{bh}$=9.4).
This is consistent  with  self-regulation of the
growth of the central SMBH, when the end of of the
accretion flow begins and the activity of the galactic
nucleus falls off. The $M_{rad}^{bh}(z)$ diagram  displays
comparatively narrow scatter, and should be preferred
for use in estimating galactic black-hole masses.

\noindent
{\small
{\bf ACKNOWLEDGMENTS}

This research has made use of the NASA/IPAC Extragalactic Database (NED),
which is operated by the Jet Propulsion Laboratory, California Institute of
Technology, under contract with the National Aero-nautics and Space
Administration. We have also used the CATS database of radio astronomy
catalogs \cite{ver2} and the FADPS system for radio-astronomical data reduction
(http://sed.sao.ru/~vo/fadps\_e.html.) \cite{ver8},\cite{ver9}. We are deeply grateful to
 R.D. Dagkesamanskii for valuable comments on the manuscript. This work was
 supported by the Program of State Support for Leading Scientific Schools of
 the Russian Federation (School of S.M. Khaikin) and the Russian Foundation
for Basic Research (projects 09-02-00298, 09-02-92659-IND, 08-02-00486). OVV
 also thanks the "Dinasty" Funding for Non-commercial Programmes for
partial support.

\end{document}